\documentclass[a4paper,fleqn]{cas-sc}

\usepackage[T1]{fontenc}
\usepackage[utf8]{inputenc}

\usepackage[numbers]{natbib}
\usepackage{mathtools}
\usepackage[scientific-notation=true]{siunitx}

\usepackage{framed} 
\usepackage{multicol} 

\usepackage{arydshln}
\usepackage{eurosym}

\makeatletter
\def\@seccntformat#1{\@ifundefined{#1@cntformat}%
   {\csname the#1\endcsname\quad}
   {\csname #1@cntformat\endcsname}
}
\makeatother
 
\usepackage{nomencl} 
\usepackage[boxruled]{algorithm2e} 
\makenomenclature
\renewcommand*\nompreamble{\begin{multicols}{2}}
\renewcommand*\nompostamble{\end{multicols}}

\renewcommand\nomgroup[1]{%
  \item[\bfseries
  \ifstrequal{#1}{A}{Abbreviations}{%
  \ifstrequal{#1}{B}{Basic definitions}{%
  \ifstrequal{#1}{V}{Variables}{%
  \ifstrequal{#1}{P}{Parameter}{%
  \ifstrequal{#1}{S}{Sets}{%
  \ifstrequal{#1}{F}{Functions and mappings}}}}}}%
]}

\begin{document}
\let\WriteBookmarks\relax
\def\floatpagepagefraction{1}
\def\textpagefraction{.001}
\shorttitle{}
\shortauthors{Accounting for spatiality}

\title [mode = title]{Accounting for spatiality of renewables and storage in transmission planning}

\author[1,2]{Leonard Göke}[type=editor,auid=000,bioid=1]
\cormark[1]
\cortext[cor1]{Corresponding author.}
 \ead{lgo@wip.tu-berlin.de}
\author[1,2]{Mario Kendziorski}[auid=000,bioid=1]
\author[2]{Claudia Kemfert}[auid=000,bioid=1]
\author[1,2]{Christian von Hirschhausen}[auid=000,bioid=1]

\address[1]{Berlin University of Technology, Workgroup for Infrastructure Policy (WIP), 10623 Berlin, Germany.}

\address[2]{Department of Energy, Transportation, Environment, German Institute for Economic Research (DIW Berlin),10117 Berlin, Germany.}

\begin{abstract}
The current governance process to plan the German energy system omits two options to substitute grid expansion: First, placing renewables closer to demand instead of where site conditions are best. Second, utilizing storage instead of additional transmission infrastructure to prevent grid congestion.

In the paper, we apply a comprehensive capacity expansion model based on the AnyMOD modelling framework to compare the status quo to alternative planning approaches for a fully renewable energy system. To represent spatiality and fluctuations of renewables, the German electricity sector is modelled with great spatio-temporal detail of 32 NUTS2 regions and hourly time-steps. In addition to the German electricity sector, analysis also accounts for exchange of energy with the rest of Europe and demand for electricity and electricity-based fuels, like hydrogen or synthetic gases, from the industry, transport, and heat sector.

The results reveal that a first-best solution can be well approximated if the current planning approach also considered storage for congestion management. Placing renewables different has no significant effect in our case, because the available potential must be exploited almost entirely leaving little room for optimization. Furthermore, a sensitivity on the first-best scenario prohibiting additional transmission lines entirely suggests that grid expansion can be substituted at tolerable costs.

\end{abstract}



\maketitle

\section{Introduction} \label{1}

To decarbonize the energy system, the primary supply of energy has to shift towards renewables like wind and solar. As a result, also the topology of power generation shifts and new options for grid planning arise. 

First, cost and potential of wind or solar greatly depend on location and, compared to fossil sources, capacities of individual plants are an order of magnitude smaller \citep{Pfenninger2014}. This imposes a trade-off on their deployment: either place plants where site conditions are best and rely on the grid to bring electricity to consumers or – to reduce the need for transmission infrastructure – place plants close to demand. 

Second, with increasing shares of wind and solar, matching intermittent generation with demand increasingly requires storage systems \citep{Schill2020}. If placed the right way, storage systems can be charged while the grid is underutilized and discharged when the grid is under stress to relieve congestion, making storage a substitute for grid expansion.

In Germany, planning transmission infrastructure is the responsibility of transmission system operators (TSO). In a continuous process the four TSOs, under regulation of the Federal Network Agency, develop scenarios for the next 15 years of power supply and use these scenarios to identify impending congestion and outages. Since Germany, a single zonal market, pursues a "copper-plate", meaning free flow of electricity within the country, it is the TSOs' task to prevent any congestion and enable market-based dispatch of all generators plus commercial exchanges with neighboring markets. Therefore, planning is focused on optimizing operation or expanding the transmission grid. Only in extreme situations or as a temporary measures to manage congestion until other projects are completed, TSOs adjust the market-based dispatch ex-post, referred to as redispatch \citep{Weber2017}.

In addition, the outlined process does not account for the two options to substitute grid infrastructure in renewable systems: placing renewables closer to demand and storage systems. Investment into generation capacities is private and driven by a single zonal market and a support scheme for renewables that is largely independent of location. Consequently, sites selected for renewables do not reflect the spatiality of demand or bottlenecks of the transmission grid. In the past this lead to a concentration of investment in the north contributing to congestion within the German market zone. For storage systems the situation is similar, the market design provides no incentive for regional investments and TSOs do not include them in the planning process. Regulation in other European countries is similar, although smaller market zones often provide better incentives for regional investments \citep{Weber2013}.

\citet{Grimm2016} and \citet{Kemfert2016} investigate how including redispatch in the planning framework and not just as a temporary measure impacts grid expansion. Both papers base their analysis on the same TSO projections for 2035, but apply different models \citep{NEP2015}. The multi-stage equilibrium model in \citeauthor{Grimm2016} relies on a stylized grid representation, but accounts for the different objectives of TSOs, private investors and the central planner. The optimization model in \citeauthor{Kemfert2016} on the other hand is limited to the central planner, but represents the power grid with greater detail instead. Both papers find that deviating from the zonal market dispatch increases social welfare and is able to substitute 57 percent of planned transmission lines according to \citeauthor{Grimm2016}, or 48 percent according to \citeauthor{Kemfert2016}, respectively. In addition, \citeauthor{Grimm2016} point out that in a first-best case where investment into generation considers grid constraints as well, the required transmission lines are reduced by two thirds. Using a very disaggregated model, \citet{drechsler2017} also find that the location of renewable energies has a clear impact on transmission requirements.

Following up on these findings, this paper investigates how including redispatch and the placement of generation and storage systems impacts system planning. In contrast to the sources above, we do not base our analysis on current energy scenarios by the TSOs, but on an own scenario that models a fully renewable energy system in Germany and Europe. This system is characterized by intermittent renewables, a consequent dependence on storage, and new demands for electricity outside the power sector. Thus, it fundamentally differs from the system analyzed in previous research.

The applied model is introduced in section \ref{2}, followed by comparative scenarios and the underlying data assumptions in section \ref{25}. The results obtained on this basis are discussed in section \ref{3}, before the a summary of key findings, policy implications and an outlook on future work follows in section \ref{4}.

\section{Applied modeling framework} \label{2}

Quantification of different planning processes follows a two-step procedure based on a techno-economic optimization model of the German energy system using the AnyMOD framework \citep{Goeke2020a, Goeke2020b}. The model chooses from a range of technologies that generate, convert, or store energy carriers to efficiently satisfy an exogenous demand.

Eqs. \ref{eq:1a} to \ref{eq:1h} provide a highly stylized version of the model formulation. To differentiate them, variables are written in capital and parameters in lower-case letters. According to the energy balance in Eq. \ref{eq:1b}, the sum of generation $Gen_{t,i,c}$, storage input $St^{in}_{t,i,c}$ and storage output $St^{out}_{t,i,c}$ over all technologies $i$ has to match demand given by the parameter $dem_{t,i,c}$ at each time-step $t$ and for each energy carrier $c$. The following storage balance connects storage in- and output with the storage level $St^{size}_{t,i}$ at each time-step $t$ for each storage technology. Eqs. \ref{eq:1d} to \ref{eq:1f} enforce capacity constraints on storage in- and output, storage levels and generation ensuring production does not exceed the capacity $Capa_{i}$. For generation, capacity constraints include a capacity factor $cf_{t,i}$ that specifies the share of capacity available for generation at time-step $t$. Finally, the objective function Eq. \ref{eq:1a} is composed of total investment costs $InvCost$ computed from capacities and specific investment costs $invCost_i$ in Eq. \ref{eq:1g} and total variable costs $VarCost$ computed from generation $Gen_t$ and specific variable costs $varCost$ in Eq. \ref{eq:1h}. For a full description of the underlying optimization model, that also includes the representation of different regions and how they can exchange energy carriers, see \citet{Goeke2020a}.  

\begin{subequations}
\begin{alignat}{4}
\text{min} & \; \; \hspace{0.94em} InvCost + \sum_{c \in C} VarCost_c & & & \label{eq:1a} \\
\text{s.t.} & \; \; \hspace{0.85em} \sum_{i \in I} Gen_{t,i,c} + St^{out}_{t,i,c} - St^{in}_{t,i,c} & \; \;  = \; \; & dem_{t,i,c} & \;  \; \forall t \in T, c \in C \hspace{0.18em} \label{eq:1b} \\ 
& \; \; \hspace{0.85em} St^{size}_{t-1,i}  + \sum_{c \in C}  St^{in}_{t,i,c} - St^{out}_{t,i,c} & \; \;  = \; \;  & St^{size}_{t,i} & \;  \; \forall t \in T, i \in I_{st} \label{eq:1c} \\ 
& \; \; \hspace{0.85em} \sum_{c \in C} Gen_{t,i,c} & \; \; \leq \; \; & cf_{t,i} \cdot Capa^{gen}_{i}  & \; \; \forall t \in T, i \in I \hspace{0.48em} \label{eq:1d} \\  
& \; \; \hspace{0.85em} \sum_{c \in C} St^{out}_{t,i,c} + St^{in}_{t,i,c}  & \; \; \leq \; \; & Capa^{st}_{i} & \; \; \forall t \in T, i \in I  \hspace{0.5em}  \label{eq:1e} \\ 
& \; \; \hspace{0.85em} St^{size}_{t,i} & \; \; \leq \; \; & Capa^{size}_i & \; \;  \forall t \in T \hspace{3.0em}  \label{eq:1f} \\
& \; \; \hspace{0.85em} \sum_{\forall i \in I } Capa_{i} \cdot  invCost_i & \; \; = \; \; &  InvCost  & \; \; \label{eq:1g} \\ 
& \; \; \sum_{\forall t \in T, \, i \in I} Gen_{t,i,c} \cdot  varCost & \; \; = \; \; &  VarCost  & \; \;  \label{eq:1h} 
\end{alignat}
\end{subequations}

\begin{figure}
	\centering
		\includegraphics[scale=.45]{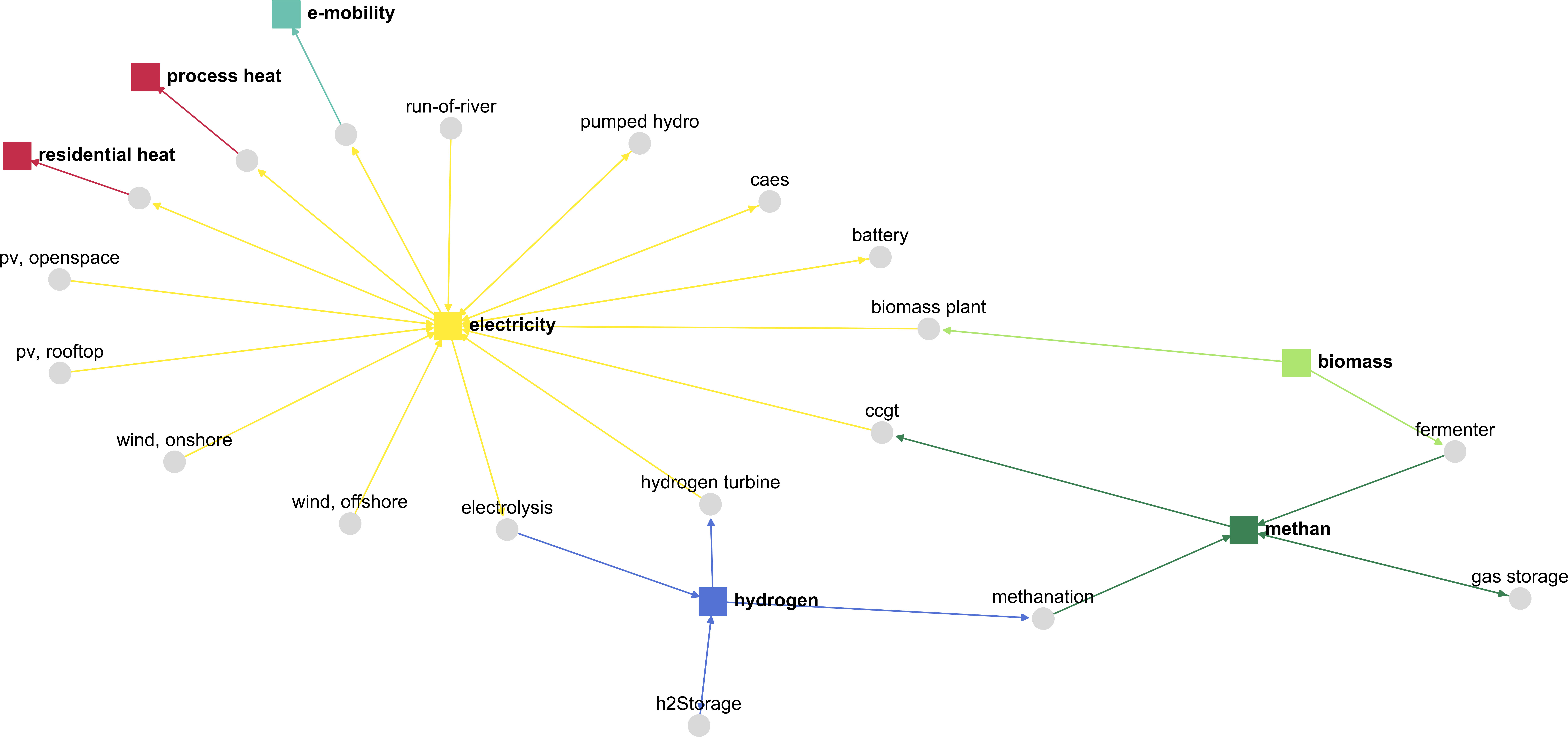}
	\caption{Graph of model elements}
	\label{fig:1}
\end{figure}

In Figure \ref{fig:1} all considered technologies, depicted as gray circles, and their interaction with energy carriers, depicted as colored squares, are visualized. Entering edges of technologies refer to their input carriers; outgoing edges relate to outputs. For example, the biomass plant uses biomass as an input to generate electricity. Storage technologies, like pumped hydro or compressed air energy storage (CAES), have an entering and an outgoing edge to represent charging and discharging. 

Due to its pivotal role for renewable systems, the model's focus is on electricity. For long-term storage of electricity, the analysis includes hydrogen and synthetic methane. Setting an exogenous demand for these carriers also captures the demand for synthetic fuels outside of the power sector, for example in aviation. Beyond that, representation of other sectors is limited to their electricity demand induced by sector integration. These demands are treated separately using the carriers "residential heat" for hot water and space heat, "process heat" for industrial heating, and "e-mobility" for electric vehicles. Demand from these sectors is exogenous since the model does not include deployment of technologies outside the power sector.
 
The demand for each carrier has to be met by the various technologies for each considered time-step and region whereby time-steps and regions can vary by energy carrier.\footnote{For a detailed description of how this feature is achieved see \citet{Goeke2020a}.} For electricity, the model applies an hourly temporal resolution to capture the fluctuating nature of intermittent renewables. Hydrogen and synthetic gas are balanced daily since they are less sensitive to short-term imbalances. Electric mobility uses a daily resolution, too, assuming vehicle charging is flexible. Lastly, residential and process heat apply a 4-hour resolution to account for the thermal inertia of buildings and load shifting potentials in the industry.

\begin{figure}
	\centering
		\includegraphics[scale=.55]{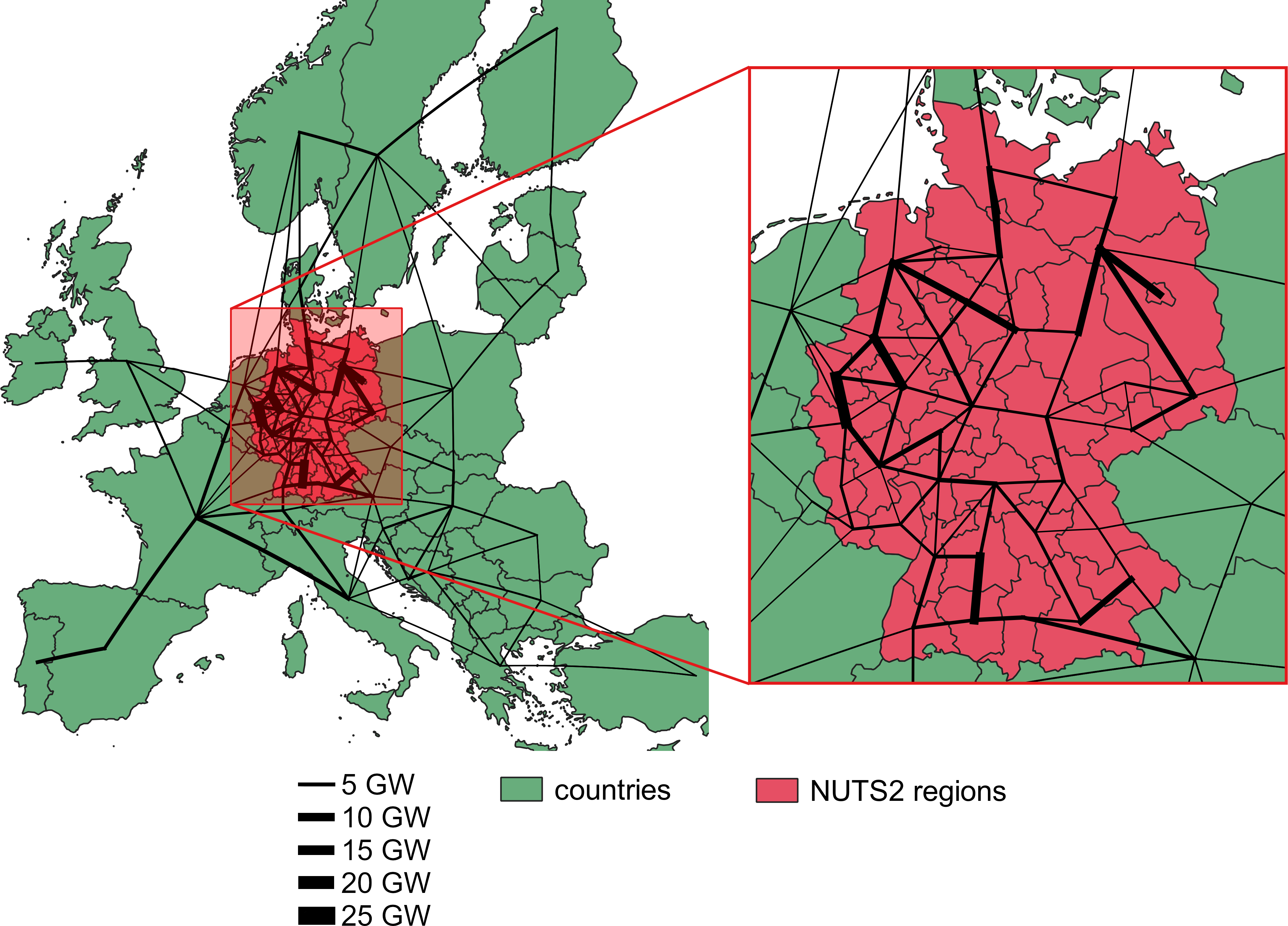}
	\caption{Overview of regions including pre-existing electricity grid, Sources: \citet{lkdeu2017,entsoe}}
\label{fig:2}
\end{figure}

The spatial resolution is uniform for all energy carriers but varies by scenario for reasons that will be elaborated on in the following section. Figure \ref{fig:2} provides an overview of all regions. These include 29 regions for European countries and 38 NUTS2 regions for Germany, which are modelled separately since our research question focuses on spatial effects and requires great regional detail.

Furthermore, the model allows for regular trading: Electricity, hydrogen and synthetic gases can be exchanged between regions, given the required grid infrastructure. Investment and dispatch for this infrastructure is, analogously to technologies, calculated by the model. Since the paper focuses on Germany, other European countries are only included to account for cross-border trade of energy. Therefore, technology and grid capacities for these countries are exogenous and the model only decides on their dispatch.

The high-spatio temporal detail for Germany paired with a representation of the European energy system and impact of sector integration on the system, is a unique feature of the model enabled by the AnyMOD framework. For deciding on investment and dispatch of technology and grid capacities, the model considers investment, operating, and dispatch costs to find the least-cost solution to satisfy the given demand. So, mathematically our approach is a linear minimization of system costs, which, since demand is exogenous and therefore assumed to be inelastic, is equivalent to welfare maximization. The model is limited to a single year and omits the transformation from today to a renewable system. Also, exchange of electricity neglects loop flows and how line expansion affects transmission losses. These simplifying assumptions are necessary to keep the computational complexity manageable.

\section{Scenarios and data} \label{25}

\subsection{Considered scenarios} 

\begin{figure}
	\centering
		\includegraphics[scale=.55]{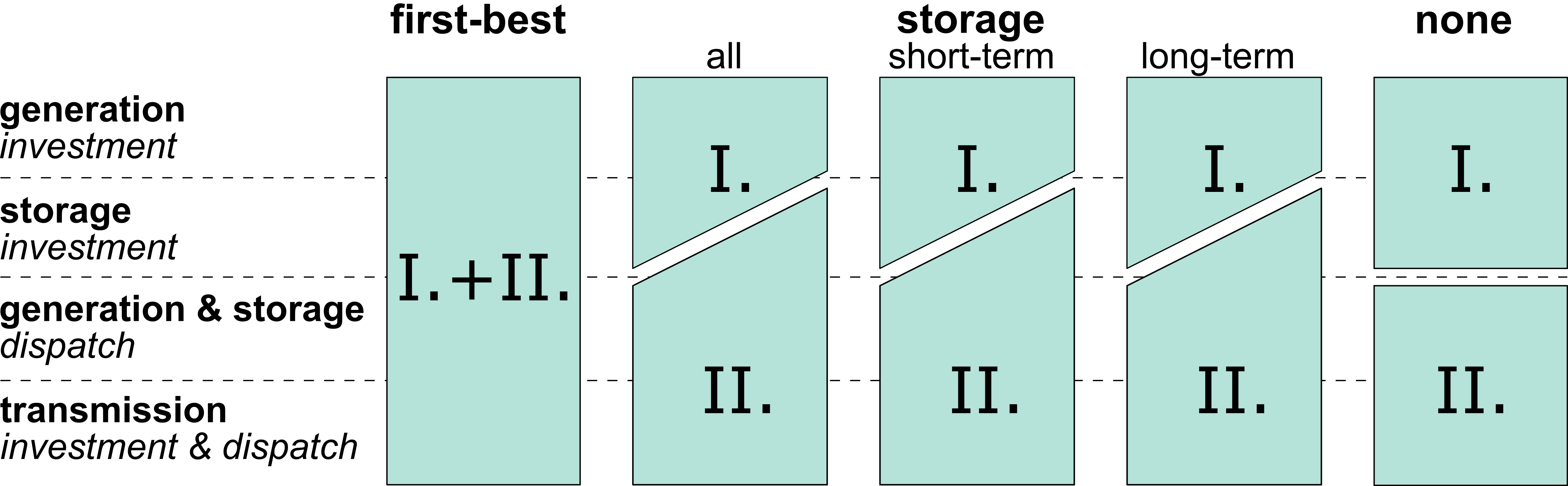}
	\caption{Overview of considered scenarios}
\label{fig:3}
\end{figure}

Analysis of the different planning processes builds on several scenarios summarized by Figure \ref{fig:3}. These scenarios differ regarding the sequence in which investment and dispatch decisions are determined. 

The first-best case on the very left only deploys the model once. Investment and dispatch of generation, storage, and transmission are all determined simultaneously for each of the 38 German NUTS2 regions. As a result, the trade-offs between grid expansion and placing generation differently, using storage systems, or deviating from a market-based dispatch are all internalized by the model. The scenario thus corresponds to a social welfare optimum. Note that this setting is currently not practiced, as it would require a major change to current regulation. A policy frequently proposed in the dedicated literature to achieve this optimum is nodal pricing \citep{Harvey2000}. To gain insight on the general importance of transmission, analysis also includes a sensitivity of the first-best without any transmission expansion at all.

In all other scenarios, investment and dispatch of generation, storage, and transmission is not determined simultaneously but sequential. The first step computes investment in generation and storage technologies, ignoring all grid constraints and assuming a free flow of electricity within Germany. Accordingly, results correspond to a market-based dispatch with a single German zone. The second step introduces the grid to determine investment into transmission, but fixes technology investment depending on the scenario. Since in the absence of grid constraints the model is indifferent where to place storage systems, these are distributed proportionally to renewable generation across the 38 regions. This is plausible given the assumed absence of regional prices, because investors have an incentive to place renewables and storage at the same sights to decrease costs for construction and grid access. If transmission losses incurred in the second step render the problem unsolvable, because demand cannot be fully met, the entire process is repeated with a correspondingly increased demand in the first step. 

Since the sequential scenarios separate investment into generation and transmission, they contrast from the first-best and represent today's planning approach. In that case, the implementation of corresponding policies likely requires less regulatory change.

The following lists all sequential scenarios detailing how they fix results from the first in the second step and what kind of planning policy is simulated this way. The list follows the order from left to right in Figure \ref{fig:3}. 
\begin{itemize}

\item \textbf{All storage}: In this scenario, dispatch decisions in the second step can deviate from the market-based dispatch determined in the first. In addition, storage investment in the first step is not binding, but serves as a lower limit instead. This means storage is considered for grid relieve in the planning process, resulting in additional storage capacities on top of market driven investments.

\item \textbf{Short-term storage}: This scenario is equivalent to "All storage", but additional storage investment is limited to short-term storage, namely battery and CAES.

\item \textbf{Long-term storage}: The scenario is again equivalent to "All storage", but now additional investment is limited to technologies for long-term storage of electricity, which are electrolysis, methanation, hydrogen plants, and gas plants.

\item \textbf{None}: In this scenario all technology investment, even storage is fixed in the second step. However, dispatch in the second step can still deviate from the market-dispatch computed in the first step.

\end{itemize}

In conclusion, only in the first-best scenario system planning considers all three substitutes for grid expansion: placement of generation, storage systems, and deviating from the zonal market dispatch. The following three scenarios consider storage and a deviating dispatch, but do not consider a different placement of generation. The last scenario only considers dispatching capacities differently.
 
\subsection{Data}

The following section summarizes the most important quantitative assumptions used in the model. To ensure consistency, as much data as possible was based on the same underlying scenario of a renewable European energy system, the "Societal Commitment" scenario developed in the openENTRANCE project \citep{Auer2020}. For comprehensive information on all inputs see the link in the supplementary material.

\subsubsection{Supply} \label{231}

For the German NUTS regions, generation and storage capacities are determined according to the outlined scenarios based on investment and operating costs \citep{OSMOSE,Auer2020,FraunhoferEE}. To account for cross-border trade, the other European countries are included in these scenarios as well, but their generation and storage capacities are fixed to not distort results. These capacities are instead computed in a preceding step using the same input data, but reducing Germany to a single node. For the sensitivity of the first-best case without grid expansion, this preceding step is carried out without any expansion of the European transmission grid.

Capacity limits and factors of renewables for the other European countries are based on \citeauthor{Auer2020}. Capacity factors from the German NUTS regions are extracted from \href{www.renewables.ninja}{renewables.ninja} \citep{ninja1,ninja2}. An input not provided anywhere in the literature are capacity limits of wind and photovoltaic (pv) broken down by German NUTS2 regions. Therefore, these assumptions were derived based on publicly available sources specifically for this study. To ensure consistency with the rest of input data, summed limits for Germany corresponds to \citet{Auer2020}. 

First, highly resolved satellite data for land use provides the urban, sub-urban, agricultural, and forested area in each NUTS2 region \citep{Corine}. Other literature gives the share for those areas that are typically suited for wind an solar \citep{Nahmacher2014,Bodis2019}. According to the product of area size and share suited for renewables, the total limit is distributed across all urban, sub-urban, agricultural, and forested areas.

Next, site quality for each of these area is extracted from geodata on average full-load hours for wind and pv \citep{solar,wind}. To derive renewable limits graded by quality in each NUTS2 region, areas are clustered into different groups based on site quality.  Capacity factors for each group are derived by scaling the original time-series according to site quality, but keeping the total energy potential of each NUTS region unchanged.

\begin{figure}
	\centering
	\includegraphics[scale=.45]{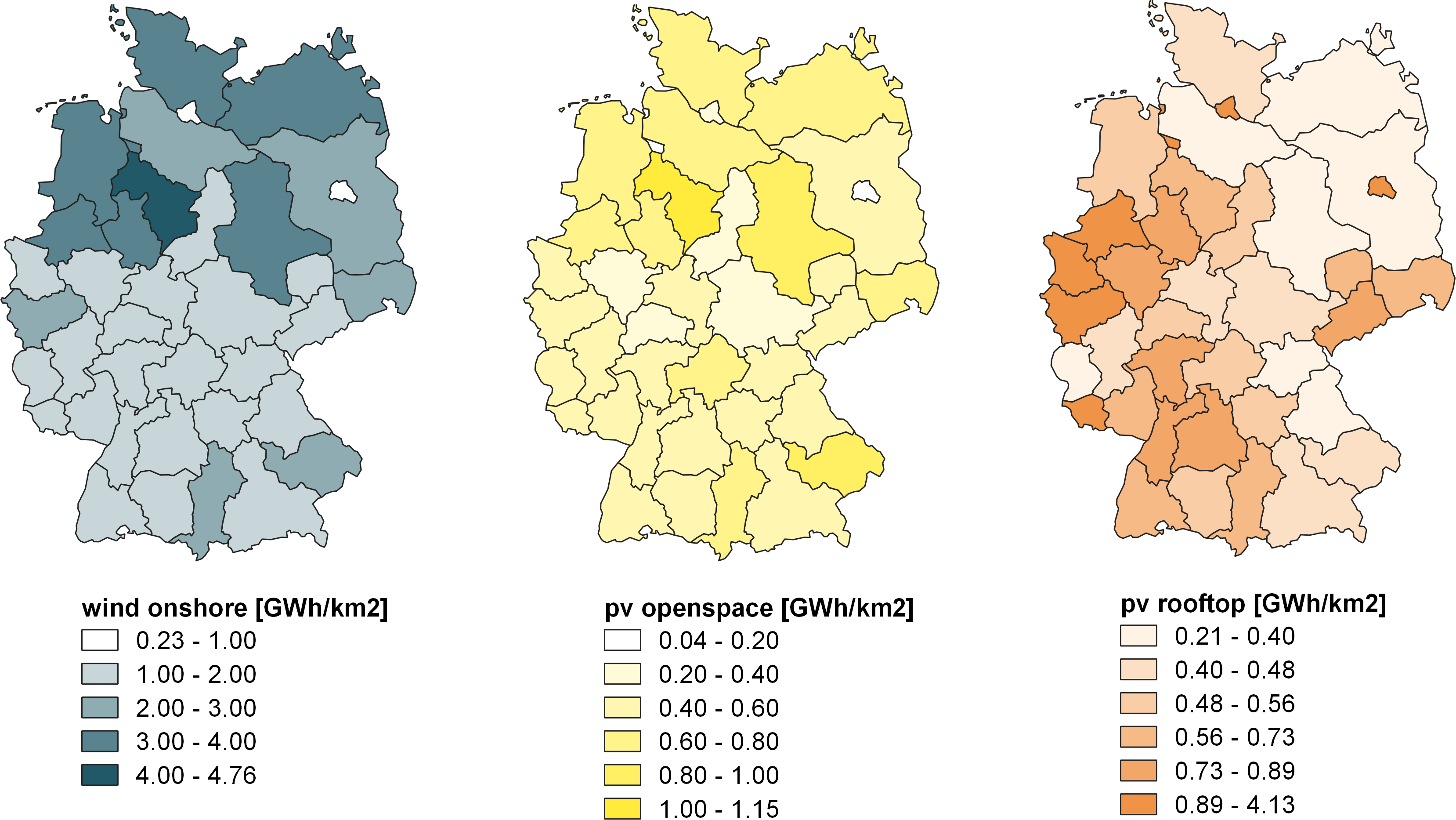}
	\caption{Comparison of energy potential per area by technology, Source: own calculations}
	\label{fig:4}
\end{figure}

Fig. \ref{fig:4} shows the resulting energy potential per area for onshore wind, openspace pv and rooftop pv. Potential for onshore wind and openspace pv is based on agricultural and forested areas and, thus, potential is highest in the least populated regions. Potential for roofop pv on the contrary relates to urban and sub-urban areas, which means potential concentration in densely populated NUTS regions, in particular cities.

\begin{figure}
	\centering
		\includegraphics[scale=.3]{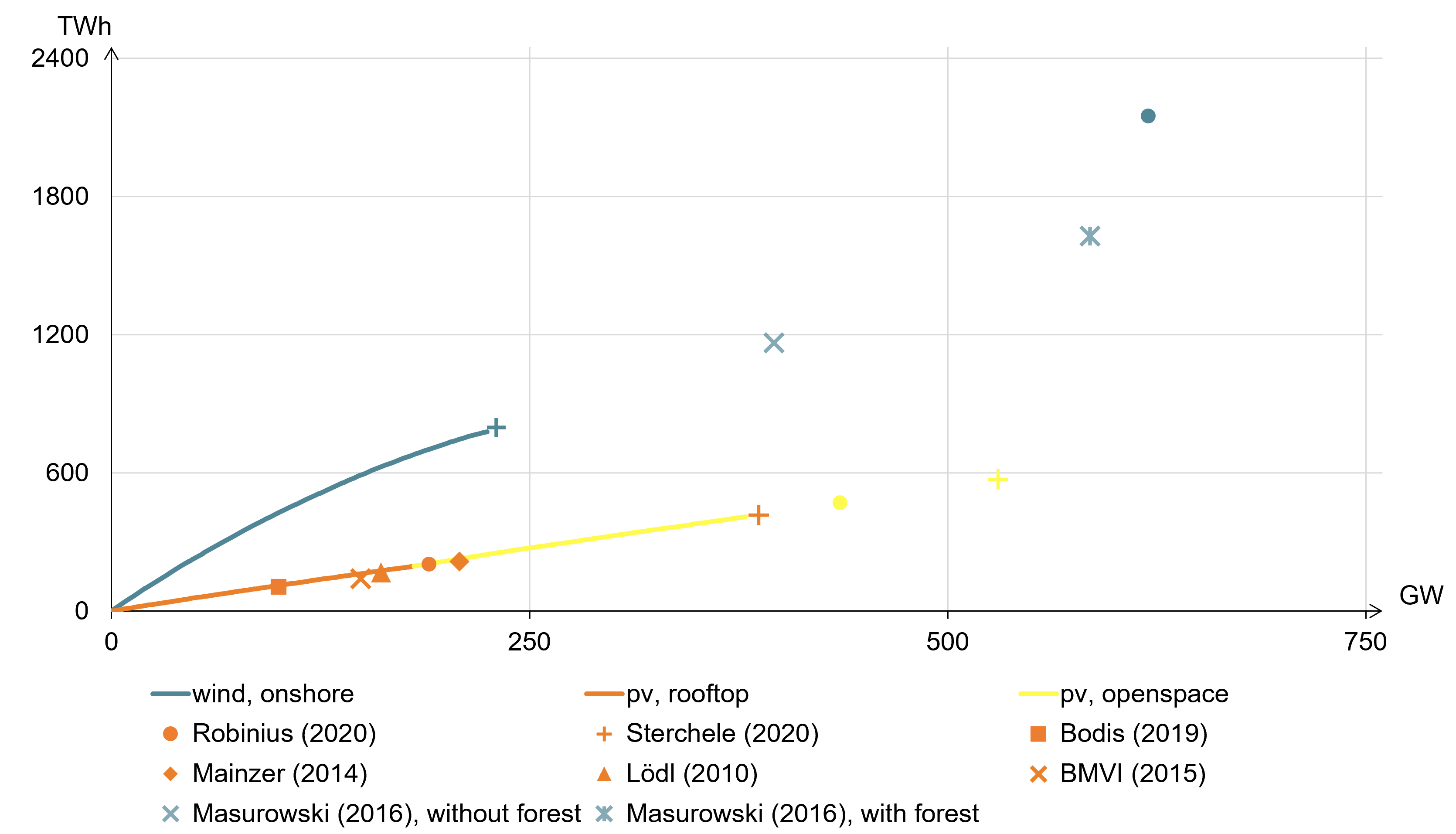}
	\caption{Renewable potentials for Germany compared to other sources (openspace cumulative to rooftop)}
	\label{fig:5}
\end{figure}

To provide some context Fig. \ref{fig:5} compares potentials used in this paper to other literature. The derived capacity limits are sorted by full-load hours, aggregated, and plotted against energy quantities. Accordingly, the decreasing slope of these lines represents the declining site quality when the share of exploited potential increases. Other sources are represented as points. Wherever these only specified a capacity limit, plotting assumed the same full-load hours as in our data. For onshore wind the assumed potential is at the lower end of values found in the literature, whereas assumptions for pv are largely in the middle of the observed range \citep{Sterchele2020,Robinius2020,Bodis2019,Mainzer2014,Lodl2010,BMVI2015,Masurowski2016}.

The set potential for offshore wind amounts to 70 GW with 5,100 full-load hours that strongly decrease due to wake effects as soon as installed capacities exceed 50 GW \citep{Agora2020}. The potential is distributed across NUTS regions currently connected to offshore wind parks.

\subsubsection{Demand}

\begin{figure}
	\centering
		\includegraphics[scale=.4]{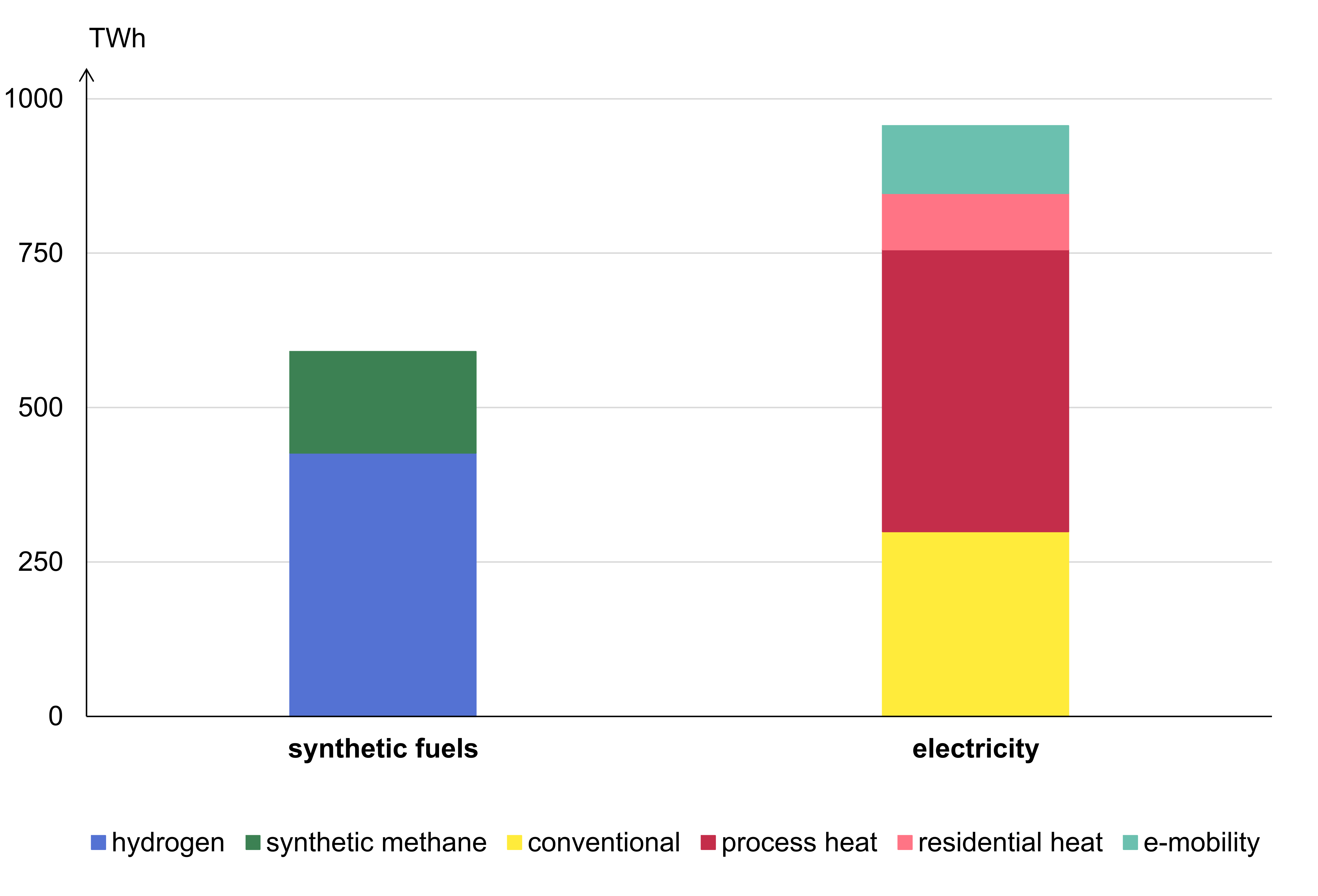}
	\caption{Demand for synthetic fuels and electricity in Germany, Source: \citet{Auer2020}}
	\label{fig:6}
\end{figure}

Given the importance of sector integration, analysis of renewable systems must consider all sectors, but our model only covers synthetic fuels and electricity explicitly. Therefore, heating and transport are implicitly included by adding the demand for synthetic fuels and electricity that decarbonization of these sectors requires \citep{Auer2020}. Fig. \ref{fig:6} provides the resulting demand for Germany; magnitude and structure are similar for other countries.

The data distinguishes between two different synthetic fuels: synthetic methane and hydrogen. According to \citeauthor{Auer2020}, synthetic methane is exclusively used to provide process heat for industrial processes. Hydrogen is also used for industrial processes, but to a small extent also for residential heating. The majority of hydrogen demand, namely 60 percent, stems from freight transport. Also, a small share is used in aviation.

As has been explained in section \ref{2}, the model treats electricity demand from process heat, residential heat, and electric mobility separately. Process heat, in particular steam generation, constitutes the largest share of electricity demand. Shares for residential heat, mostly heat-pumps, and electric mobility are considerably smaller. Finally, demand that does not fit into any of these categories, for example household appliances, is labelled "conventional".

For Germany, national demand from \citeauthor{Auer2020} has to be distributed across the 38 NUTS2 regions modelled. For this purpose, electricity demand from process heat is distributed according to gross domestic product, residential heat according to reported heating demand, and mobility according to population \citep{eurostat2020,eurostat2021,hre2017}.

\subsubsection{Transmission}

For representation in the model, the physical transmission infrastructure is aggregated according to the covered regions. Due to the long lifetime of transmission infrastructure, the current electricity grid displayed in Fig. \ref{fig:2}, is available in the model without additional investments. For Europe, these pre-existing capacities built on TSO data on net transfer capacities and include all projects to be completed by 2025 \citep{entsoe}. Capacities between German NUTS2 regions are aggregated from a nodal dataset. Apart from electricity, the model also includes a representation of today's gas grid assuming future utilization for hydrogen \citep{lkdeu2017}. However, today's capacities were found to already exceed future needs. For this reason, transport restrictions for hydrogen are neglected within Germany.

The model represents transmission infrastructure as net transfer capacities and consequently simplifies their dispatch to a transport problem neglecting technical constraints. Investment costs and losses of transmission depend on the length of the aggregated lines as displayed in Fig. \ref{fig:2} and amount to 2.29 million Euro per GWkm and 5 percent per 1000 km, respectively \citep{JRC,Neumann2020a}. 

\section{Results} \label{3}

The results first focus on the first-best scenario and its sensitivity to create an understanding for the modelled system in general and for the role of the transmission infrastructure in that system in particular. This understanding is then necessary to comprehend the comparison of the first-best to the sequential planning scenarios in the second part. 

\subsection{First-best scenario}

\begin{figure}
	\centering
			\includegraphics[scale=.35]{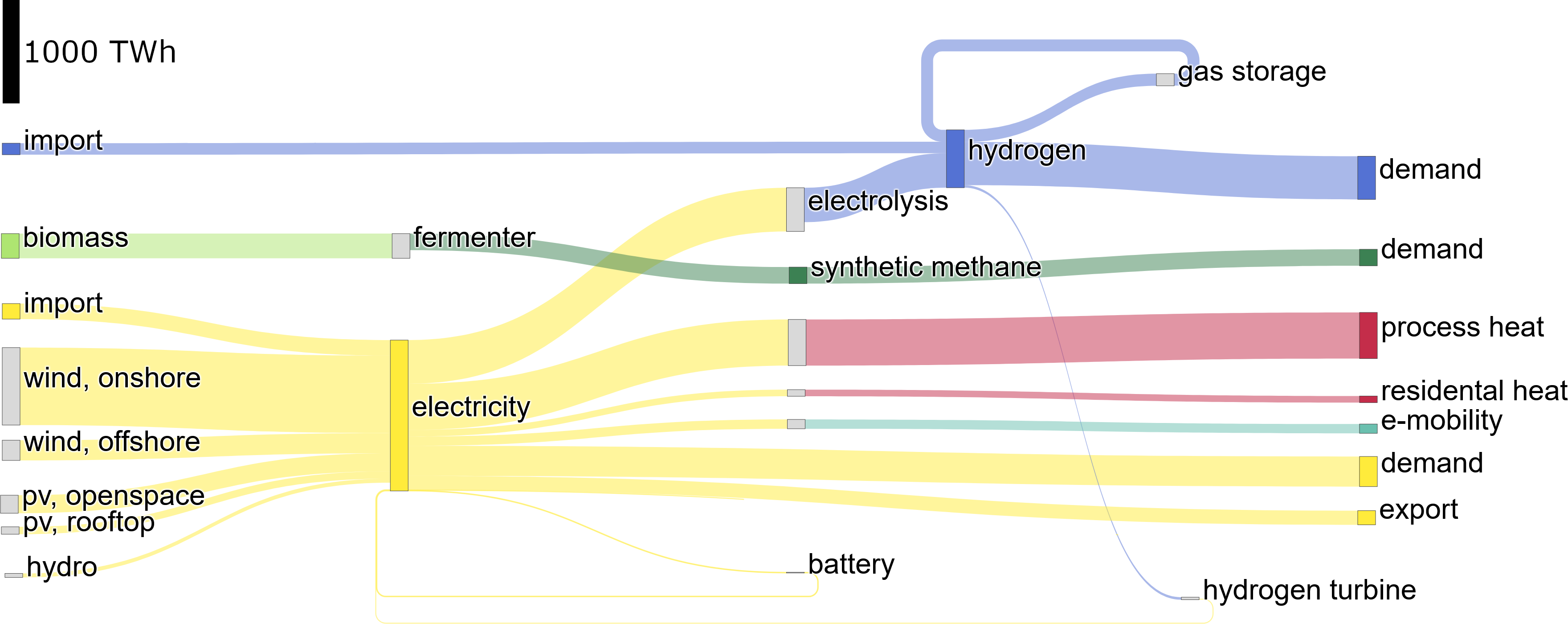}
	\caption{Energy flows in first-best scenario}
	\label{fig:7}
\end{figure}

Fig. \ref{fig:7}, the quantitative counterpart to Fig. \ref{fig:1}, shows the energy flows for Germany that result from solving the model for the first-best scenario. Energy flows for the other scenarios differ of course but show no fundamental differences. Total electricity demand amounts to 1350 TWh and is covered by 766 TWh of generation from wind onshore, 200 TWh from wind offshore, 178 TWh from openspace pv, 74 TWh from rooftop pv and lastly 39 TWh from hydro, which includes run-of-river and reservoirs. With regard to Fig. \ref{fig:7}, this means all renewables technologies except rooftop pv fully exploit their energy potential. Since sector integration makes up most of the demand and is assumed to be flexible within certain limits, storage systems only play a relatively minor role. Batteries provide 18 TWh of electricity and 16 TWh of electricity are generated from stored hydrogen, while the larger share of hydrogen satisfies the exogenous hydrogen demand. In addition, 114 TWh of hydrogen are imported from other European countries. Electricity is both imported and exported leading to an import surplus of 13 TWh. The demand for synthetic methane is entirely met through biomass, independent from the rest of the energy system.

\begin{figure}
	\centering
			\includegraphics[scale=.42]{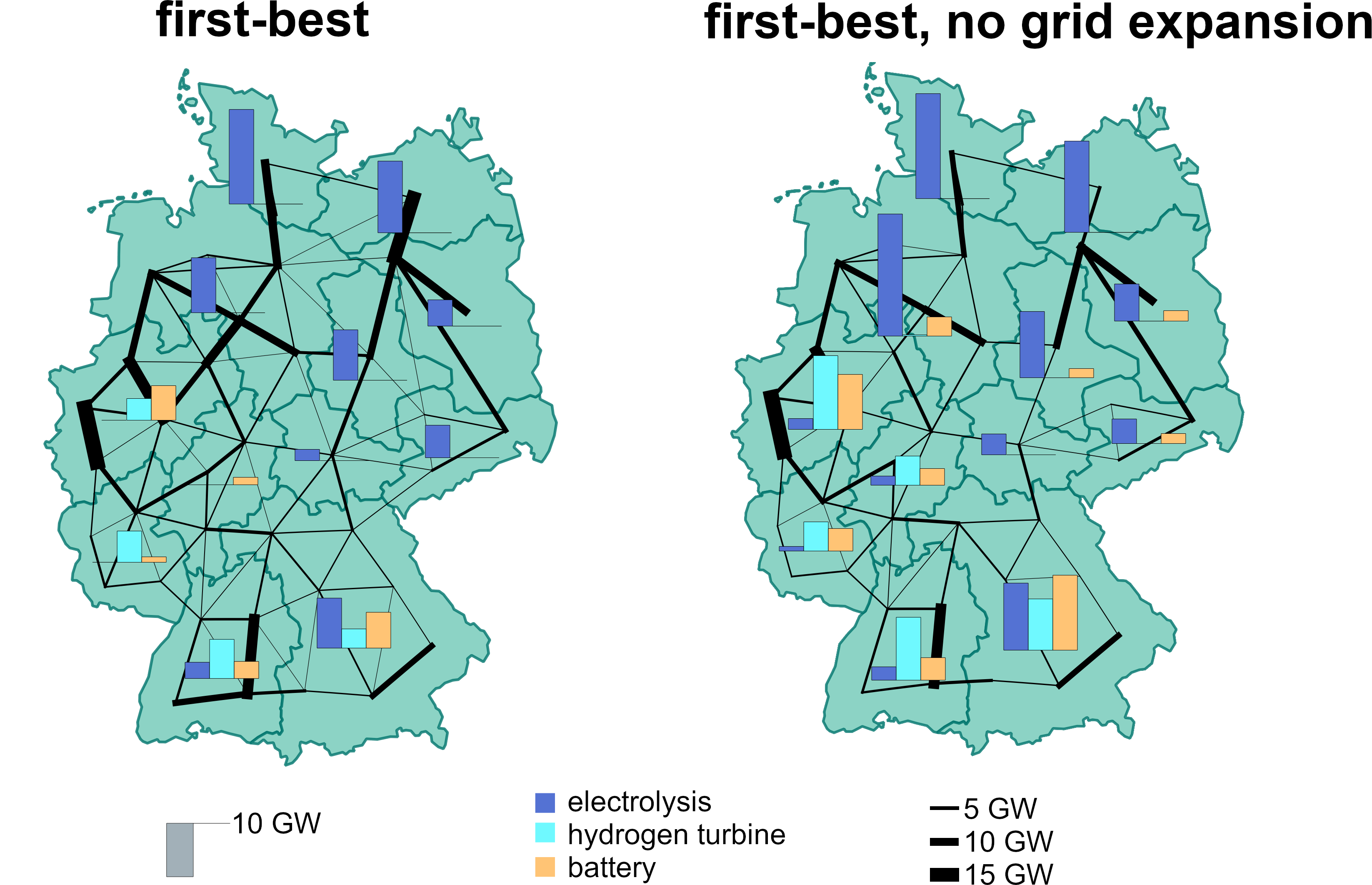}
	\caption{Storage capacities and grid expansion for first-best}
	\label{fig:8}
\end{figure}

If the first-best is solved without grid expansion, generation from rooftop pv increases by 50 TWh, but plant capacities do not shift and just increase in regions with unexploited potential. Instead, long- and short-term storage substitute for grid expansion; generation from hydrogen turbines increases to 55 TWh, output from batteries to 27 TWh. This substitution can also be observed when mapping grid and storage capacities as done in Fig. \ref{fig:8}.\footnote{For better illustration the figure shows technology capacities aggregated by NUTS1 regions. Small NUTS1 regions, like city states, were assigned to their nearest neighbor.} If grid expansion is disabled, less capacity is available to transport electricity from regions in the north with large potential to regions in the south and southwest with highest demand. In return, capacities of hydrogen turbines and batteries increase substantially in these regions.

Since grid expansion occurs in the first-best solution, disabling it will rise system costs, with higher investment costs for generation and storage overcompensating the decrease of transmission costs. For Germany system costs for Germany rise by 2.5 percent, but the more meaningful comparison of European system costs shows an increase of 4.5 percent. This seems plausible given \citet{Neumann2020b} observe a 10 percent increase of system costs when limiting capacities to today's grid when modeling a renewable European power system, but note that consideration of sector integration is likely to reduce this difference. Such reduction could be explained by the added flexibility and high utilization of renewables potential from sector integration.

\subsection{Sequential scenarios compared to first-best}

To compare the first-best with today's planning framework, grid expansion and system costs in Germany are compared for different scenarios in Table \ref{tab:comp}. For a sensible benchmark of grid investment, expansion of each line is multiplied with its length and totalled. In comparison, the pre-existing grid that is available without additional investments amounts to 39,653 GWkm. 

As expected, system costs and grid expansion are smallest for the first-best scenario. If grid expansion is determined after generation and storage, but considers all storage technologies as a substitute, results only show a slight increase in grid expansion. However, in the scenarios that only consider short-term storage or no storage at all, expansion increases substantially doubling the capacity of the pre-existing grid. If the second planning step only allows for additional long-term storage, viz. electrolyzers and hydrogen turbines, grid capacities increase by 50 percent. Long-term storage presumably has a more pronounced effect than short-term storage because strictly speaking it does not only allow for additional storage, but also to shift demand to some extent since electrolyzers also have to satisfy an exogenous hydrogen demand and hydrogen can be freely transported within Germany. 

Difference in system costs are closely correlated with grid expansion. Overall, the largest proportion of costs, about 77 percent in the first-best scenario, is incurred by generation. The next factor is transmission costs, accounting for 12 percent of costs, followed by long- and short-storage with 9 and 2 percent, respectively. Since transmission only makes up a relatively small proportion of system costs, they are affected less severely by the different scenarios. Still, system costs increase by up to 8 percent if grid expansion does not consider long-term storage, which is higher than in the first-best case without grid expansion. 

\begin{table}[ht]
\centering
\begin{tabular}[t]{lcc}
\hline
&grid expansion \textit{\small [GWkm]} &system costs \textit{\small [Bil. \euro]}\\
\hline
first-best & 8,734 & 51.69 \\ \hdashline
all storage & 9,781 & 51.94 \\
short-term storage & 40,640 & 55.58 \\
long-term storage & 17,274 & 52.75 \\
none & 40,654 & 55.59 \\
\hline
\end{tabular}
\caption{Key benchmarks of scenarios compared} 
\label{tab:comp}%
\end{table}%

Again, the scenarios show little difference with regard to the placement of generation, because capacity limits are almost fully exploited to satisfy demand. Difference are most significant when comparing the first-best to the "none" scenario. To compensate for lower capacity factors from placing renewables closer to demand, the first-best installs 13 GW more rooftop pv and 5 GW more onshore wind. 

\begin{figure}
	\centering
			\includegraphics[scale=.42]{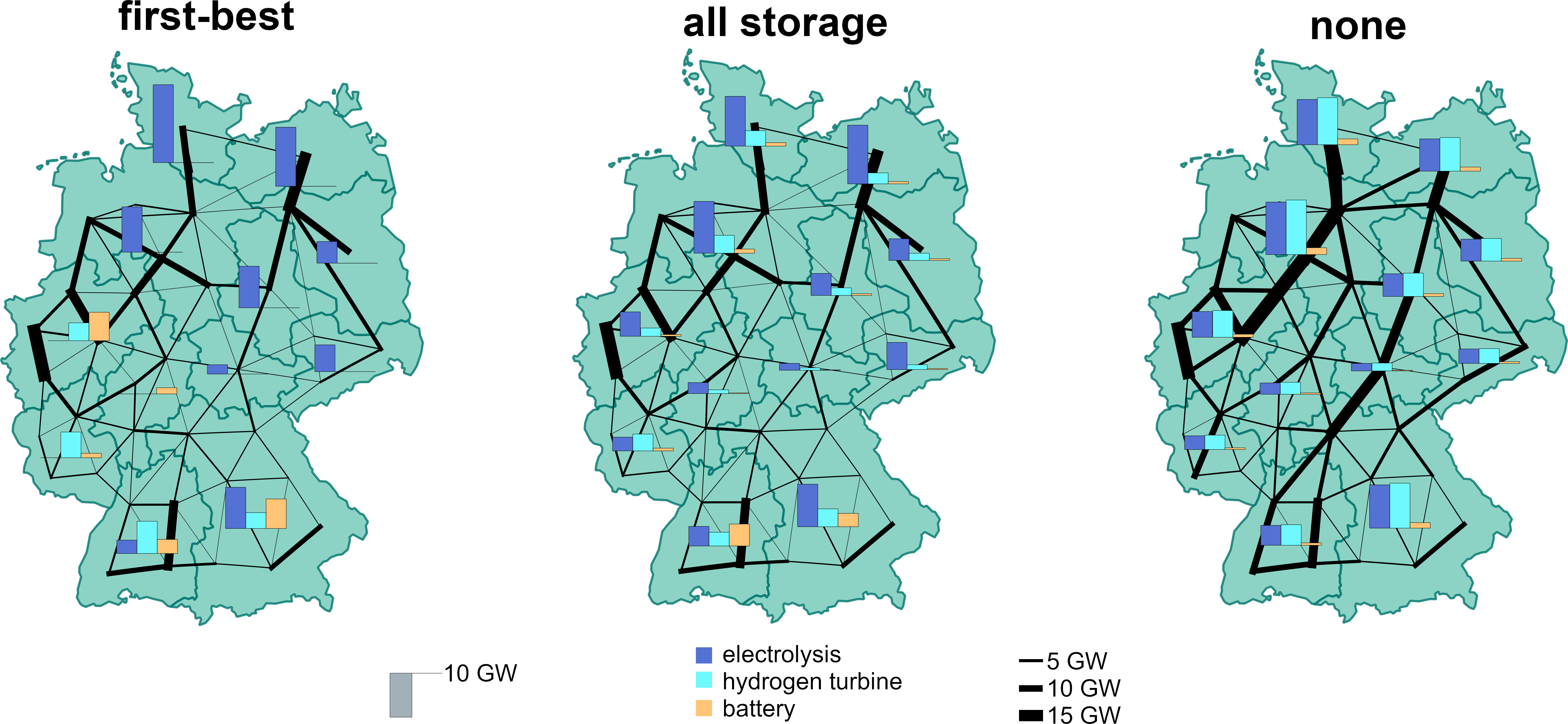}
	\caption{Storage capacities and grid expansion compared to first-best}
	\label{fig:9}
\end{figure}

The trade-off between grid and storage is again visualized in Fig. \ref{fig:9}. Between the first-best and the scenario with storage as a substitute ("all storage"), no differences are visible with regard to transmission capacities. For storage investment on the other hand, there are significant differences. In the first-best, hydrogen turbines and batteries are exclusively located in importing regions with high demand. In the storage scenario, capacities are concentrated in these regions too, but since here only the second step of investment considers the grid, all regions have some capacity. The map on the right shows capacities, if no substitute for grid expansion is considered. Here, storage capacities are evenly distributed and grid capacities, especially from exporting regions in the North to importing regions in the South, are considerably higher. Also, average utilization of transmission capacities decreases to 909 full-load hours, compared to 1,300 in the first-best.

\section{Conclusions} \label{4} 

In this paper we applied a capacity expansion model to investigate substitutes for transmission infrastructure in renewable energy systems and how use of these substitutes depends on the underlying planning approach. The model is applied to the German power sector but takes detailed account of sector integration and cross-border exchange.

Results show that consideration of storage, in particular long-term storage, for congestion management greatly decreases grid investment and thus also system costs. If this option is enabled by a first-best setting that optimizes investment and dispatch of generation, storage, and transmission simultaneously, or, more similar to today's planning process, independent from generation investment in a subsequent step together with transmission investment, does not make significant difference. Findings also suggest that grid expansion can be substituted largely by storage placed in high-demand regions, causing a 4.5 percent increase in system costs.

On the other hand, very small effects on welfare and grid expansion were observed from considering grid constraints when placing renewables. These results are partly driven by the characteristics of the underlying scenario for the entire energy system used in this study. As section \ref{231} shows, the assumed capacity limits for renewables, in particular for wind, are at the lower end of literature values. Therefore, the available potential for renewables is almost fully exploited to satisfy demand leaving little room to optimize their placement.

Similar to findings on redispatch in \citet{Grimm2016}, results indicate that transmission planning can substantially benefit from modifications to the current policy framework. The first-best solution requiring invasive changes, like the introduction of nodal pricing, can be well approximated, if planning considers storage as a substitute for grid expansion. Conceivable instruments to this end are an obligation for TSOs to consider storage investments or a split of the Germany market into a north-east and south-west price zone to create incentives for private storage investment. In both cases, consumers in exporting regions benefit at the expense of importing regions, either in the form of lower grid charges or smaller market prices. Apart from that, the relatively small sensitivity of grid expansion on system costs suggests that an exclusive focus on costs is too narrow. Given the public opposition transmission faced in the past, a bearable increase in system costs might be preferable to high levels of grid expansion.

The applied capacity expansion framework AnyMOD captures all important features of renewable energy systems: intermittent renewables, importance of storage, increased demand and added flexibility from sector integration, as well as cross-border trade of electricity. In return, some economic and technical aspects relevant for transmission planning had to be neglected, which may limit the significance of our results. On the economic side, the approach omits path dependencies by focusing on a single year and abstracts from the different agents involved in the planning process, which hinders the recommendation of more specific policy instruments. On the technical side, restrictions of operating power grids, like physical power flows or n-1 security, were omitted. Also, although dividing Germany into 38 different regions is comparatively detailed, it does not compare to the 500 nodes of the actual transmission grid represented in power system models. To cover these aspects, future research needs to adapt economic and technical models to the characteristics of renewable energy systems.

\section*{Acknowledgements}
The research leading to these results has received funding from the German Federal Ministry for Economic Affairs and Energy via the project "MODEZEEN" (grant number FKZ 03EI1019D). The authors thank Karen Pittel and others for valuable feedback on this work received through the seminar on the "Economics of Fossil Fuel Phase-out" and the MultiplEE workshop 2021. Also, we thank Richard Weinhold for valuable comments on an earlier draft of this paper, and Urban Persson for providing NUTS level data on heat demand from the Heat Roadmap Europe project on request. The usual disclaimer applies.

\section*{Supplementary material}

All computations in this paper were run with version 0.1.6 of the modelling framework AnyMOD.jl (\url{https://github.com/leonardgoeke/AnyMOD.jl}). All scripts and data files to replicate results are available on Zenodo (\url{https://doi.org/10.5281/zenodo.4569880}). The upload also includes additional information on the input parameters used.

\bibliographystyle{elsarticle-num-names}
\bibliography{cas-refs}

\end{document}